\documentclass[reprint, amsmath,amssymb,aps,twocolumn]{revtex4-1} 
\usepackage{graphicx}
\usepackage{dcolumn}
\usepackage{bm}
\usepackage{color}      
\usepackage{setspace}
\usepackage{subfigure}
\usepackage[plainpages=false, pdfpagelabels,
    pdfpagemode=none,
    pdfstartview=FitH,
    colorlinks=true,
    linkcolor=blue,
    citecolor=blue,
    a4paper=true,
    pagebackref=false,
    pdfstartpage=1,
    pdfauthor={F.Mirjani},
    pdfsubject={Notes},
    pdftitle={},
]{hyperref}
\begin{document}
\preprint{APS/123-QED}
\title{DFT-based many-body analysis of electron transport through molecules}
\author{F.~Mirjani and J.~M.~Thijssen}
\affiliation{Kavli Institute of NanoScience, Delft University of Technology, Lorentzweg 1, 2628 CJ Delft, The Netherlands }
\date{\today}
\begin{abstract} 
We present a method which uses density functional theory (DFT) to treat transport through a single molecule connected to two conducting leads for the weak and intermediate coupling. This case is not accessible to standard non-equilibrium Green's function (NEGF) calculations. Our method is based on a mapping of the Hamiltonian on the molecule to a limited set of many-body eigenstates. This generates a many-body Hamiltonian with parameters obtained from ground state L(S)DA-DFT calculations. We then calculate the transport using many-body Green's function theory. We compare our results with existing density matrix renormalization group (DMRG) calculations for spinless and for spin-1/2 fermion chains and find good agreement.  
\begin{description}
\item[PACS numbers]
\end{description}
\end{abstract}
\pacs{Valid PACS appear here}
\maketitle
\section{Introduction}
The rapid development in the field of electrical transport through molecules and quantum dots has induced a considerable effort to investigate the physical mechanisms behind it. For a conceptual understanding of these phenomena it is indispensable to develop methods involving a minimal number of approximations. Different schemes have been used for the calculation of the conductance of such systems. The most popular ones are ab initio methods based on density functional theory (DFT) \cite{KohnSham} in connection with the non-equilibrium Green's function (NEGF) formalism which has been successfully used for understanding coherent transport through molecules in the strong coupling or off-resonant regime \cite{NEGF-DFT1, NEGF-DFT2, Smeagol}. Also DFT is known to yield good results for ground state calculations. 
Although it has been argued by Stefanucci and Almbladh \cite{Almbladh} that appropriate time dependent functional should also give good results for transport, the currently available functionals are not satisfactory for transport in the weak coupling regime where the Coulomb interaction between the electrons dominates their dynamics \cite{PeterFerdinand}. Using the available ground states DFT functionals in particular gives a poor description of ionization, addition and excitation energies and these states play an important role in transport.

To illustrate the failure of DFT, we consider the usual case of non-ferromagnetic leads which always yields a solution in which spin up and down have the same occupation. Transport through a single level through which one or two electrons can flow, is described by a single-level Anderson
type model. The most general form of the density matrix (restricted to the transport level) during transport, is given by:
\begin{equation}
\rho=\vert 0\rangle \langle0\vert+a\vert \uparrow \rangle \langle\uparrow \vert+a\vert \downarrow \rangle \langle\downarrow \vert+b\vert \uparrow \downarrow \rangle\langle\uparrow \downarrow \vert
\end{equation}
Local density approximation (LDA) in DFT yields for this case a restricted solution, which does not distinguish between the last three terms and it is well known it cannot produce the correct step-like behaviour of the current as a function of voltage \cite{Stadler}. Instead, for each level, the current rises gradually with bias up to a maximum value. Part of the shortcoming of DFT(LDA) may be corrected for by adding a self-interaction correction (SIC) which first leads to a plateau corresponding to a single conduction channel before it steps to a next plateau at maximum current corresponding to the two conducting channels \cite{Datta2, SIC}. Although the SIC method is an improvement over the standard DFT, it still gives wrong results for the current value through a single plateau: the SIC method predicts the current value of the plateau to be half of the maximum current while, as we will show in Sec \ref{result1}, it should be $2/3$ of the maximum current.\\ 
These shortcomings have induced the development of different methods for weak coupling regime. As an example, the many-body effects that are not captured by DFT-NEGF can be obtained by the GW approximation method, which however is very time-consuming \cite{Thygesen}.\\ 
Combining DFT with rate equations can be used to describe the electron transport in the weak coupling regime \cite{Beenakker, Seldenthuis1} but this technique requires fit parameters and cannot show the broadening of the isolated levels due to the coupling (except for the temperature broadening). However, DFT is a powerful means to calculate the total ground state energies and this leads us to exploit this advantage of DFT in this regime. Thus our purpose is to present a technique relying on the combination of DFT and many-body NEGF approach which deals with transport in the weak coupling regime. Our method combines local spin density approximation (LSDA) for different numbers of electrons with many-body Green's functions (GF) to calculate the transport through a molecule, weakly connected to two non-interacting leads. 
We illustrate our method using an interacting hopping chain for particles with and without spin. The latter case allows for a comparison with density matrix renormalization group (DMRG) calculations \cite{dmrg1, dmrg2}. We not only obtain excellent values for addition and ionization energies, but also good agreement of the location and the line shapes of these resonance levels when comparing with results based on DMRG method. The line shapes are the result of the coupling between the states on the molecule to the leads, which we also calculate using our DFT states. It is envisaged that the method of the paper will be useful within ab initio quantum chemistry calculations for electron transport.\\
The organization of this paper is as follows. In section \ref{sectionModel} the model for spinless and spin-1/2 fermions is defined and then our method is explained. The results for the single level inside or near the bias window are discussed in Sec \ref{result1}. Then the results for the more complicated case with two levels inside the bias window are presented in Sec \ref{result2}. The conclusions in Sec \ref{conclu} briefly summarize our ideas. The appendices include further details concerning Bethe-Ansatz solution for spinless fermions (\ref{appx1}), L(S)DA-DFT for the Hubbard model (\ref{appx2}) and calculating the transport through a Coulomb island (\ref{appx3}).\\
In this paper, we use the term `level' to indicate a chemical potential corresponding to an energy resonance on the molecule.
\section{Model and method}
\label{sectionModel}
\subsection{spinless fermions}
\label{sectionspinlessfermions}
\begin{figure}
\includegraphics[scale=0.7]{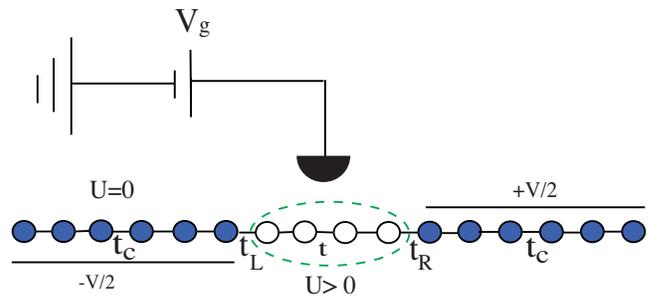}
\caption{\label{fig:model}A short Hubbard chain connected to two non-interacting leads.}
\end{figure}

The systems studied here consist of a small region where Coulomb interactions are present, weakly coupled to two non-interacting, semi-infinite leads (see Fig.~\ref{fig:model}).
The interacting region contains one or several quantum dots in series. The Hamiltonian of the entire system is
\begin{equation}
H=H_{\text {leads}}+H_{\text{coupling}}+H_{\text{molecule}}
\end{equation}
The Hamiltonian for spinless fermions with interaction reads \cite{Alexander}:
\begin{eqnarray}
&\displaystyle H_{\text{molecule}} =-t \sum_{i=1}^{N_L-1} [d^\dagger_{i} d_{i+1 }+h.c.] \nonumber \\
&\displaystyle + U \sum_{i=1}^{N_L-1} (n_i-\frac{1}{2})(n_{i+1}-\frac{1}{2}) + \epsilon \sum_{i=1}^{N_L}  d^{\dagger}_{i } d_{i }
\label{HM1}
\end{eqnarray}	
where $n_i=d^\dagger_{i} d_{i }$ and $N_L$ is the length of the interacting chain. The parameter $t$ represents the hopping rate and $U$ describes the inter-site Coulomb interaction. The creation and annihilation operators, $d^\dagger_{i}$ and $d_{i}$ acting on site $i$ satisfy the usual anticommutation relations. In addition, the external gate potential, $V_g$, can be applied to the interacting region which is included in the energy $\epsilon$.\\
For the noninteracting leads, 
\begin{equation}
H_{\text{leads}}=-\sum_{\eta=L,R} t_c \sum_{i=1}^{N_L-1} [c^\dagger_{i,\eta} c_{i+1,\eta}+h.c.]
\end{equation}
where $t_c$ is the hopping term in the contact part, and the label $\eta=L,R$ for left $(L)$ and right $(R)$ lead. The eigenstates are $\psi^{\sigma}_n=e^{\pm ika n}$ with energy \cite{Datta}
\begin{equation}
E=E_0-2t_c\cos{ka}
\label{-2tccoska}
\end{equation}
where 
\begin{equation}
e^{ika}=-q\pm\sqrt{q^2-1} \qquad , \qquad
q=\frac{E-E_0}{2t_c}
\end{equation}
We take $a\equiv 1$. In addition, the bias voltage can be applied to the contacts.\\
The coupling Hamiltonian reads
\begin{equation}
H_{\text{coupling}}= \sum_{\substack{\eta=L,R\\ j\in \text{molecule}}} [t_{\eta}c^\dagger_{i,\eta} d_{j}+h.c.]
\end{equation}
The Hamiltonian for the central part, $H_{\text{molecule}}$, can be solved exactly using the Bethe-Ansatz solution \cite{Orbach}. For such a system, Takahashi \cite{Takahashi} gives the equations which should be solved for the density $n$. This is briefly explained in Appendix \ref{appx1}. 
\subsection{spin-1/2 fermions}
Adding the spin as an extra degree of freedom, $H_{\text{leads}}$ and $H_{\text{coupling}}$ are similar to the described Hamiltonians for the spinless case but a spin index $\sigma=\uparrow,\downarrow$ is added to the creation and annihilation operators.
\begin{equation}
H_{\text{leads}}=-\sum_{\eta=L,R} t_c \sum_{\sigma} \sum_ {i=1}^{N_L-1} [c^\dagger_{i,\eta,\sigma} c_{i+1,\eta,\sigma}+h.c.]
\end{equation}
The Hamiltonian of the interacting region reads
\begin{eqnarray}
&\displaystyle H_{\text{molecule}} =-t \sum_{\sigma} \sum_ {i=1}^{N_L-1} [d^\dagger_{i,\sigma} d_{i+1, \sigma}+h.c.] \nonumber \\
&\displaystyle + U \sum_{i=1}^{N_L} d^{\dagger}_{i \uparrow} d_{i \uparrow} d^{\dagger}_{i \downarrow}  d_{i \downarrow}+\epsilon \sum_{\sigma} \sum_{ i=1}^{N_L}  d^{\dagger}_{i \sigma} d_{i \sigma}
\label{HM2}
\end{eqnarray}
In this case, the Coulomb energy is on-site (as two particles may now occupy the same site) and we parametrize it again by $U$.
\subsection{Method}
\label{Method-label}
\begin{figure}
\includegraphics[scale=0.42]{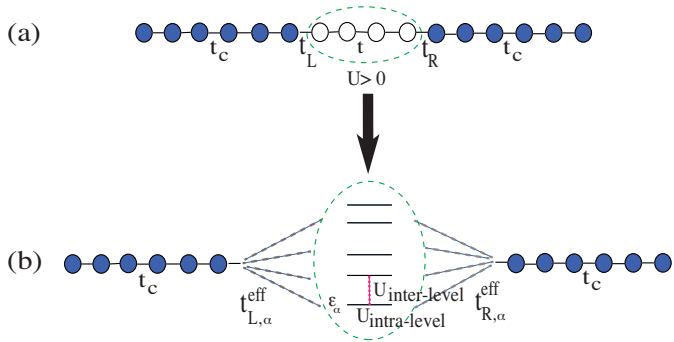}
\caption{\label{fig:mapmodel}The states of the interacting chain in (a) are mapped onto a single dot which contains several interacting levels (b). The coupling between these levels and the leads are defined as $t^{\text{eff}}$. The new model includes the intra-level Coulomb interaction and the inter-level interaction between the states.}
\end{figure}

Our method for calculating fermion transport through the interacting chains starts by expressing the molecular Hamiltonian in terms of its (many-body) exact eigenstates $\vert S\rangle$:
\begin{equation}
H_{\text{molecule}} = \sum_{S} \vert S \rangle E_S \langle S \vert
\label{SS}
\end{equation}
Because of the two-body character of the Coulomb potential, we can formulate this Hamiltonian in terms of creation and annihilation operators for (spin-) orbitals $\vert \alpha \rangle$ with a Coulomb interaction:
\begin{equation}
H_{\text{molecule}} = \sum_{\alpha} \varepsilon_{\alpha} d^{\dagger}_{\alpha} d_{\alpha}+ \frac{1}{2} \sum_{\alpha \neq \beta} U_{\alpha \beta} d^{\dagger}_{\alpha} d_{\alpha} d^{\dagger}_{\beta} d_{\beta}
\label{HM}
\end{equation}
Note that the quantum number $\alpha$ includes the spin (for spin-1/2 particles). The eigenstates $\vert S\rangle$ are then the states $\vert {n_{\alpha}}\rangle$, where $n_{\alpha}=0,1$ represents the occupation of all (spin-) orbitals $\vert \alpha \rangle$. Note that Eq (\ref{HM}) is a reformulation of the original Hamiltonian (Eq (\ref{HM1}) or (\ref{HM2})) in terms of an interacting multi-level interacting Anderson model.
Eq (\ref{SS}) is another reformulation of these Hamiltonians. We shall use Eq (\ref{HM}) to find the specific form of the many-body eigenstates S appearing in Eq (\ref{SS}).\\
Our method is based on a mapping of the Hamiltonian on the molecule to a limited set of many-body eigenstates (Fig.~\ref{fig:mapmodel}). We first find the set of parameters $(\varepsilon_{\alpha},U_{\alpha,\beta})$ from DFT ground state calculations. For the interacting chains used in this paper, we use the LDA parameterization based on the Bethe-Anstaz solution for interacting fermion chains \cite{bethe1, Orbach, Takahashi, Capelle, Capellesame}. In the case of spin-1/2 particles, we have used an accurate LSDA parameterization, given by Fran\c{c}a, Vieira and Capelle (FVC) \cite{capelle2}.\\
We first consider particles without spin. DFT allows for calculating ground state energies for any number of particles. For a chain of $N$ sites as in Fig.~\ref{fig:mapmodel}a, this number varies between 0 and $N$, so DFT gives us $N+1$ energies. Considering the system as in figure Fig.~\ref{fig:mapmodel}b, we would need $N$ chemical potentials $\varepsilon_{\alpha}$ and $N(N-1)/2$ Coulomb interactions between these levels which adds up to $N(N+3)/2$ parameters. It is therefore clear that this parametrization is highly non-unique. The situation is different in the case of spin-1/2 particles. For particles with spin 1/2, we can vary the particle number $M$ between $0$ and $2N$ and we can vary the polarization $(M_{\uparrow}-M_{\downarrow})$ between $- \text{Min}(M,N)$ and $ \text{Min}(M,N)$ yielding $N(N+3)/2$ different ground state configurations precisely the number of parameters needed in the Fig.~\ref{fig:mapmodel}b.

For small bias voltage, the transport is dominated by a single chemical potential, corresponding to a transition from $N$ to $N+1$ particles. For this case, we can still apply our method to spinless particles.\\
We calculate transport for the system consisting of a molecule described by the Hamiltonian (\ref{HM}), coupled to the noninteracting leads. It is therefore necessary to evaluate the coupling for the different (spin) orbitals $\vert \alpha \rangle$ to the leads. We do this by projecting the original chain Hamiltonian (\ref{HM1}) or (\ref{HM2}) onto two many-body states of the isolated central region, differing by one particle. We call those states $\vert S_{N-1}\rangle$ and $\vert S_N \rangle$. $S_N$ is obtained from $S_{N-1}$ by putting a particle into level $\alpha$ which is empty in $S_{N-1}$, $\vert S_{N}\rangle \equiv \vert S_{N-1}, \alpha\rangle$. We explain the method for the spinless case. The Hamiltonian, formulated in the space spanned by $\vert S_N \rangle$ and $\vert S_{N-1}\rangle$ is 
\begin{eqnarray}
\tilde{H}=\vert S_{N-1} \rangle E_{s_{N-1}} \langle S_{N-1} \vert + \vert S_N \rangle E_{s_N} \langle S_N \vert.
\end{eqnarray}
The coupling Hamiltonian describes the hopping of a particle from or onto the left lead to the leftmost site of the central region and a similar description can be used for the right lead. We anticipate that the effective coupling of a level $\alpha$ to the leads varies with the amplitude of that state on the left- and rightmost sites respectively. We calculate this effective coupling using the projection operator:

\begin{equation}
P=\vert S_{N-1} \rangle \langle S_{N-1} \vert + \vert S_N \rangle \langle S_N \vert
\end{equation}
The process in which a particle hops from the left lead onto the leftmost site of the central region is described
by the following term of the coupling Hamiltonian (the calculation for the hopping to the right lead is similar):
\begin{equation}
\hat{H}_{\text{coupling}}=t_{L}d^{\dagger}_{1} c_L.
\end{equation}
The full Hamiltonian with the central region projected onto the subspace spanned by $S_{N-1}$ and $S_N$, contains transitions of the form $H_{\text{coupling}} = t^{\text {eff}} \vert S_N \rangle  \langle S_{N-1} \vert c_L$. Projecting $\hat{H}_{\text{coupling}}$ onto the span of $S_N$ and $S_{N-1}$ gives
\begin{eqnarray}
& \displaystyle{ \hat{P}^{\dagger} \hat{H}_{\text{coupling}} \hat{P}= t_L (\vert S_N \rangle \langle S_N \vert) d^{\dagger}_{1} c_L (\vert S_{N-1} \rangle \langle S_{N-1} \vert)} 
\end{eqnarray}
using $d^{\dagger}_1 \vert S_{N-1} \rangle= \vert 1;S_{N-1} \rangle$ leads to
 \begin{eqnarray}
& \displaystyle{ \hat{P}^{\dagger} \hat{H}_{\text{coupling}} \hat{P}= t_L \vert S_N \rangle \langle S_N \vert 1 ; S_{N-1} \rangle \langle S_{N-1} \vert c_L=}  \nonumber \\
& \displaystyle { t^{\text {eff}}_L \vert S_N \rangle \langle S_{N-1} \vert c_L }
\end{eqnarray}
which requires $\displaystyle {t^{\text {eff}}_L=t_L \langle S_{N}  \vert 1;  S_{N-1} \rangle } $. Here $L$ is the site of the left lead connected to the central region, and $\vert 1;S_{N-1} \rangle$ denotes an antisymmetrized state obtained by adding an electron on site $1$ to a central region
containing $N-1$ particles in state $\vert S_{N-1} \rangle$.


In DFT, the approximated eigenstate is given as 
\begin{equation}
\vert S_N \rangle=\frac{1}{\sqrt{N!}}\sum_{P}\eta_P \vert \varphi_{P_1}^N ... \varphi_{P_N}^N \rangle ,
\end{equation} 
i.e. a Slater determinant composed of the single-particle DFT orbitals $\varphi_{k}^N$ found within the $N$ particle ground state ($\displaystyle {\sum_P}$ is a sum over permutations and $\eta_P$ is the sign of the permutation). Defining $\vert \varphi_{N}^{N-1}\rangle\equiv \vert 1 \rangle $, $t_L^{\text {eff}}$ reduces to
\begin{equation}
t_{L,\alpha}^{\text {eff}}= t_{L} \sum_{P}\eta_P \prod_{n} \langle \varphi_{P_n}^{N-1}  \vert \varphi_{P_n}^{N} \rangle= t_{L}\times \text{det}(S)
\end{equation}
where $S$ is the "overlap matrix", $S_{kl}=\langle  \varphi_{k}^{N-1} \vert \varphi_{l}^N \rangle$ and $\alpha$ denotes the highest orbital of $S_N$. In the case of spin-1/2 particles, the calculation of the effective coupling depends on $\varphi^{\uparrow}$, $\varphi^{\downarrow}$ which leads to $t^{\text {eff}}=\displaystyle{t_{L,R}\times \text{det}(S^{\uparrow})\times\text{det}(S^{\downarrow})}$.\\
The effective coupling mainly depends on the shape of the orbitals, in particular their values on the outermost sites of the molecule. However, this shape for an electron with spin-up also depends on whether a spin-down electron occupies the level. We account for this by writing, for the coupling of a spin-up electron
\begin{equation}
{t^{\text {eff}}_{L,R}}^{\uparrow}(n_{\downarrow})=(1-n_{\downarrow}) {t^{\text {eff}}_{L,R}}^{\uparrow} (n_{\downarrow}=0) + n_{\downarrow} {t^{\text {eff}}_{L,R}}^{\uparrow}(n_{\downarrow}=1) 
\end{equation}
It turns out that the values of the coupling for the two occupations $n_{\downarrow}=0$ and $n_{\downarrow}=1$ differ only slightly (less than $2$ percent). We neglect the influence of the occupation of the other orbitals.\\
Our method for calculating the transport now consists of the following steps:(i) Calculate the ground states of the molecule for different charge states $N$ and polarizations $p=M_{\uparrow}-M_{\downarrow}$ using L(S)DA-DFT. (ii) Infer the values for $\varepsilon_{\alpha}$ and $U_{\alpha \beta}$ from these results. (iii) Calculate the effective coupling for the (spin-) orbitals $\vert \alpha \rangle$. (iv) Calculate the transport for the Hamiltonian (\ref{HM}) coupled to non-interacting leads by an $\alpha-$ dependent coupling obtained in step (iii).\\

The retarded and advanced GFs, $G^r$ and $G^a$, for the transport calculation can be derived from the equation of motion. In order to find the lesser GF, $G^{<}$, we use the Kadanoff-Baym equation
\begin{equation}
G_0^{-1} G^<= \Sigma^r G^< + \Sigma^< G^a 
\end{equation}
For details see Appendix \ref{appx3}. 
Once these GFs are known, the current can be calculated from a Landauer type of equation
\begin{equation}
I=\frac{ie}{h}\int \text{Tr}\{ \frac{\Gamma_L\Gamma_R}{\Gamma_L+\Gamma_R}(G^r-G^a)\} 
(f(\omega,\mu_L)-f(\omega,\mu_R)) \,d\omega
\end{equation}
where $\Gamma_j=i(\Sigma_j^r- \Sigma_j^{r^{\dagger}})$ and $\Sigma_j^r$ is the retarded self-energy and $f(\omega,\mu_j)$ is the Fermi distribution of lead $j$.

A few remarks are in order. In practice, we select only a limited set of many-body states, notably those whose charge additions and ionizations correspond to a chemical potential inside or near the bias window. This means that we neglect the low-lying (spin-) orbitals (that are always occupied) and the higher orbitals (that are never occupied). This enables us to treat the spinless fermion transport with low bias, even though we cannot find all the values $\varepsilon_{\alpha}$ and $U_{\alpha\beta}$ in that case.\\
Calculating the transport is a standard problem for a single orbital inside the bias window for spinless fermions. However, approximations are necessary as soon as Coulomb interactions become relevant. We follow the simplest approach, in which correlation with the leads are neglected \cite{Haug, Meir}. This means that we will not observe the Kondo resonance for spin-1/2 fermions. More elaborate schemes are possible, in particular slave-boson techniques which do take these correlations into account \cite{Slaveboson}.\\
Even within our approach, transport through a single orbital for spin-1/2 fermions is already nontrivial. For more orbitals, several schemes based on further approximations have been devised (see e.g. B. Song \textit{et al.} \cite{Bo}). We treat the full problem of spin-1/2 transport through two orbitals (four spin-orbitals) neglecting only the correlation with the leads. For details see Appendix \ref{appx3}. This already allows for 8 transport channels (7 in the case of degenerate levels). In molecular electronics, bias voltages are hardly ever high enough to observe that many states, so we do not consider larger systems.\\
A similar approach has been proposed by Yeganeh \textit{et al.} \cite{Sina} based on quantum chemistry calculations for ground states and excited states. Also a time-dependent version of LDA functional for a similar model has been used by Kurth \textit{et al.} \cite{Kurth} to investigate the transport within time-dependent DFT. Other approaches to describing transport in the weak coupling limit were based on the configuration interaction method for the central region, in combination with rate equations \cite{Datta2006-1, Datta2006-2} and with integration over scattering states constructed through a Wigner transform \cite{Delaney2004}.
\section{Results for a single level inside the bias window}
\label{result1}
In this section, we first present the results for the spinless fermions and compare with DMRG results obtained by other groups \cite{Alexander}. Then we discuss the result for spin-1/2 particles. We consider small bias, so that at most one level lies inside or near the bias window. Energies and parameters with the dimension of the energy can from now on always be assumed to be given in Volts(V) and the current and conductance units are $e/h$ and $e^2/h$ respectively.\\
\subsection{Spinless fermions}
\begin{figure}

\includegraphics[ scale=0.25]{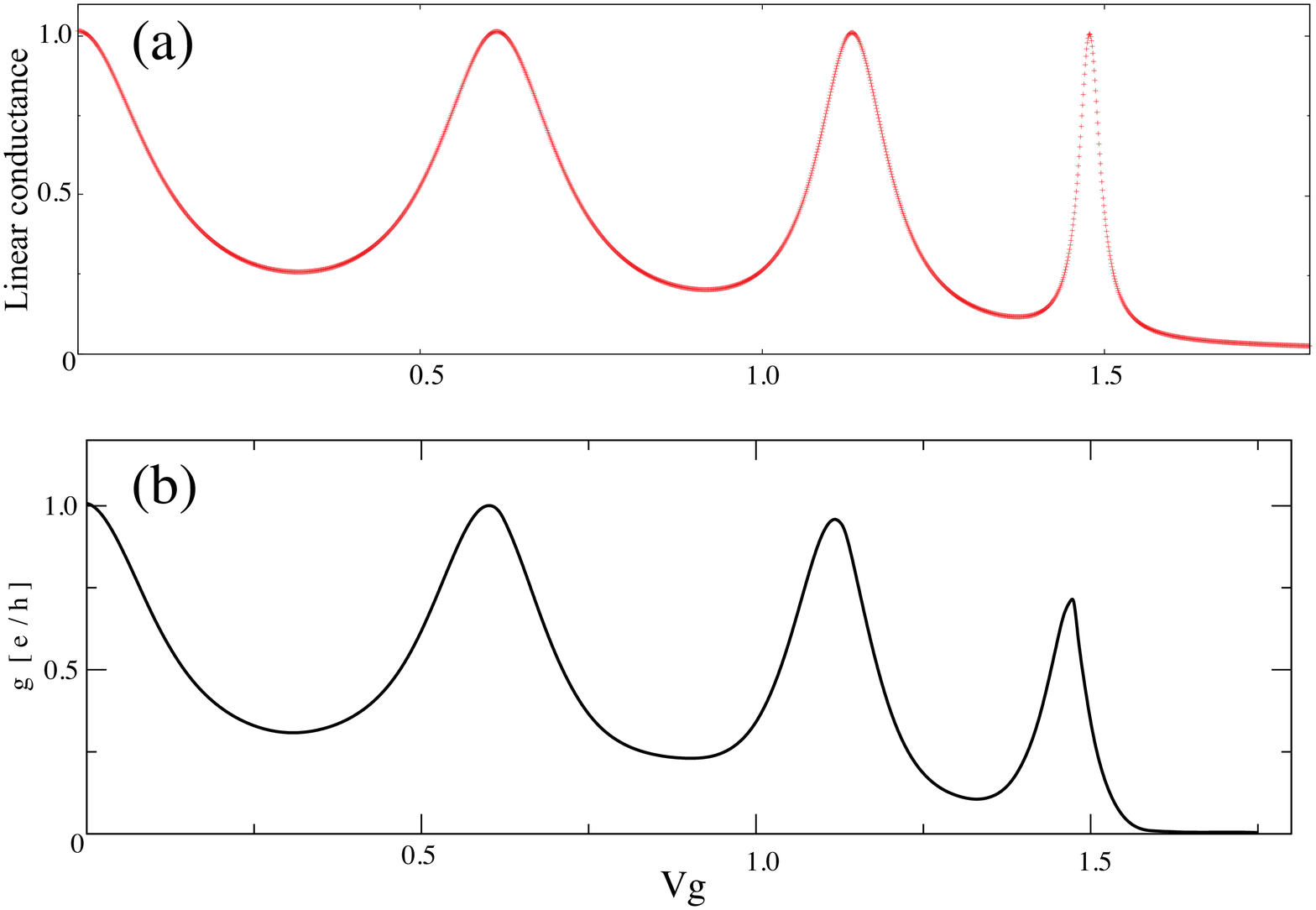}
\caption{\label{fig:noninteract.eps}linear conductance of 7 noninteracting dots, $U=0$. (a) our results and (b) the results from Ref \cite{Alexander} (Copyright Wiley-VCH Verlag GmbH and Co. KGaA. Reproduced with permission) .$V_b=2 \cdot 10^{-4}$, $t_{L,R}=0.5$, $t_c=1$, $t=0.8$.\\}

\end{figure}

\begin{figure}

\includegraphics[scale=0.45]{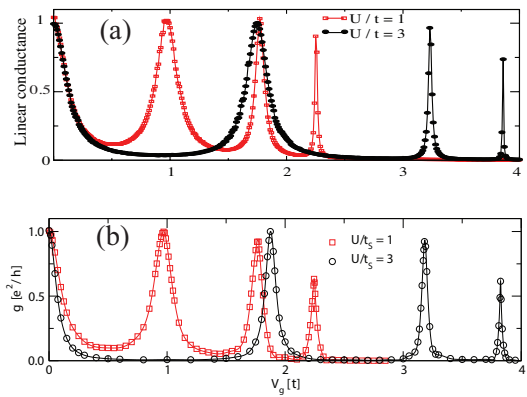}
\caption{\label{fig:interact.eps}linear conductance of 7 interacting dots for weak (squares) and strong (circles) interaction. (a) our results and (b) the results from Ref \cite{Alexander} (Reproduced with permission). $V_b=2 \cdot 10^{-4}$, $t_{L,R}=0.5$, $t_c=1$, $t=0.8$.}

\end{figure}

We first consider the results for $U=0$. The linear conductance versus the gate voltage in the case of spinless fermions is shown in Fig.~\ref{fig:noninteract.eps}a for 7 non-interacting sites compared to the Fig.~\ref{fig:noninteract.eps}b obtained from Ref \cite{Alexander} based on the DMRG method. The gate voltage is applied in order to shift different resonant levels across the narrow bias window. For this case, the peak locations are easy to get at the right position, i.e. $-2t \cos(ka)$, where $t$ is the inter-dot coupling and $k$ is the wave vector that fits on an isolated chain of 7 dots. The agreement between peak widths in the two figures shows that our way of calculating the effective coupling seems correct. The last peak is higher in our simulation than in the DMRG result, due probably to a limited number of $V_g$ values used in the latter. The linear conductance versus the gate voltage for 7 interacting sites in the cases of weak $(U/t=1)$ and strong $(U/t=3)$ interactions are shown in Fig.~\ref{fig:interact.eps}a. Since this is also in agreement with DMRG results in Fig.~\ref{fig:interact.eps}b, we conclude that our method is reliable. In both figures~\ref{fig:noninteract.eps} and \ref{fig:interact.eps}, applying the negative gate voltage, will show three peaks for the linear conductance in that region.\\
\subsection{Spin-1/2 fermions}
\begin{figure}
\includegraphics[scale=0.45]{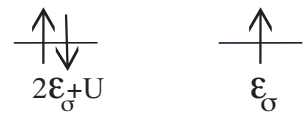}
\caption{\label{fig:esigma-u}two different situations to extract $\varepsilon_{\sigma}$ , $U$ by FVC parameterization.}
\end{figure}

\begin{figure}
\begin{center}
\includegraphics[scale=0.25]{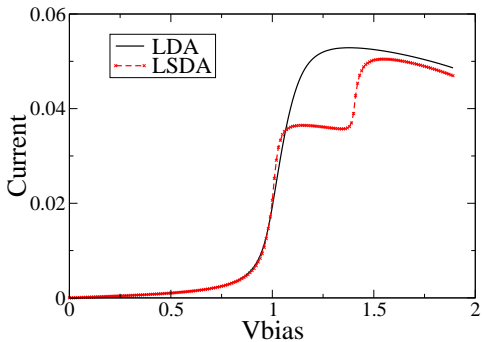}
\caption{\label{fig:ldalsda2}current through one quantum dot for LDA and for the many-body combined with LSDA (FVC parameterization) for $\mu_{R,L}=E_F \pm V/2$, $E_F=0$, $\varepsilon=0.5$, $t_{L,R}=0.1$, $t=1.0$ and $U=0.2$.\\}
\end{center}
\end{figure}

\begin{figure}
\begin{center}
\includegraphics[scale=0.25]{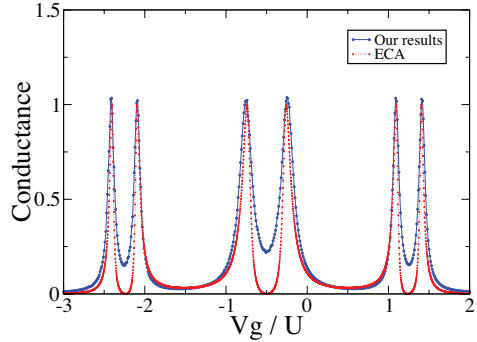}  
\caption{\label{fig:3dotscon}Conductance of three quantum dots system compared to the results of embedded-cluster approximation (ECA) of Ref \cite{Fabian}. $U=1$, $t_{L,R}=0.3$, $t_c=1$, $t=1$, $V_{bias}=0.006$.}
\end{center}
\end{figure}

For spin-1/2 particles, the ground state energies for one site containing one and two electrons have been calculated using the FVC parameterization which gives the values for $\varepsilon$ and $U$ (Fig.~\ref{fig:esigma-u}). The many-body approach described here can then be applied to calculate the current through one or several quantum dots. The result for one dot is shown in Fig.~\ref {fig:ldalsda2} for $E_F=0$, $\varepsilon=0.5$ and $U=0.2$. Our hybrid method gives two steps at the expected positions $V=2|\varepsilon-E_F|=1.0$ and $V=2|\varepsilon-E_F+U|=1.4$. We compare this calculation with the LDA-DFT-NEGF method, using the LDA parameterization by Capelle \textit{et al.} \cite{Capelle, Capellesame} (see Appendix \ref{appx2}). The LDA curve gradually increases from  $V=2|\varepsilon-E_F|$ till about $1.2 V$ where it reaches a level corresponding to transport through both channels. We see that including the spin explicitly into the transport calculation makes a substantial difference, even though on average the spin on the molecule is zero. Therefore, using a restricted exchange correlation functional is bound to give wrong results.\\
As we explained in the introduction, the SIC predicts the value of half of the maximum before it steps up to its maximum value \cite{Datta2}, while the correct value of the current in the weak coupling limit at this region is $2/3$ of the maximum current. This can be found using rate equation calculations \cite{Datta3} and it can be understood from the fact that there are two one-electron channels (corresponding to spin up and down), but only one two-electron channel.\\
The negative slope after the second step is due to the fact that the density of states in the leads is not constant. Therefore, at different biases, the leads supply a different number of electrons. Indeed, in the case of wide band limit (where self-energies are independent of energy and bias voltage) the negative slope disappears.\\
We have also compared our results for three coupled dots with DMRG in combination with the embedded-cluster approximation (ECA) of Ref \cite{Fabian} in Fig.~\ref{fig:3dotscon}. The resonance peak position is precisely in agreement with the Fig.~12c of Ref \cite{Fabian} and also the general line-shapes of the peaks are very similar to that where two central peaks are wider and four peaks at two sides are narrower than the central ones but the lowest values between peaks are different. Fig.~12c of Ref \cite{Fabian} is made assuming even numbers of electrons in the contacts and hence neglects the correlations held responsible for the Kondo resonance. Although difference with Fig.~12a of Ref \cite{Fabian}, which is believed to be the correct, seems significant, we expect them to be much less pronounced at higher bias. As it is, the agreement shows that within the approximation made in our Green's function approach, our method gives the correct prediction. An improvement which would include these correlation will be considered in future work.\\
\begin{figure}
\includegraphics[scale=0.3]{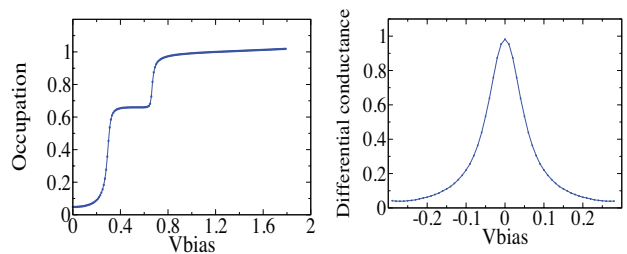}
\caption{\label{fig:5dots-total}Left: Five coupled dots system. $E_F=2.4$, $U=1.0$, $t_{L,R}=0.4$, $t_c=2$, $t=1.0$. Extracted values are $\varepsilon_{\sigma}=2.54$, $U=0.18$, $t^{\text{eff}}_{{L,R}_{N\to N+1}}=0.1$, $t^{\text{eff}}_{{L,R}_{N+1\to N+2}}=0.1$. Right: differential conductance for five quantum dots. $E_F=\varepsilon_{\sigma}=2.5479$, $U=1.0$, $t_{L,R}=0.5$, $t_c=1$. Extracted values for coupling are $t^{\text{eff}}_{{L,R}_{N\to N+1}}= t^{\text{eff}}_{{L,R}_{N+1\to N+2}}=0.143375$ and $U=0.184$. The shape of the curve is in agreement with DMRG results (Ref \cite{Alexander}).}
\end{figure}
Fig \ref{fig:5dots-total} shows the occupation of the Hubbard site $(n_{\uparrow}+n_{\downarrow})$ versus applied bias voltage for five coupled dots with $E_F=2.4$. We find $\varepsilon=2.54$ and $U=0.18$ and we see the steps at $V=2|\varepsilon-E_F|=0.14$ and $V=2|\varepsilon-E_F+U|=1.1$. The density of the level at zero bias is non-zero which is due to the broadening of lowest unoccupied molecular orbital (LUMO) which contributes in the transport. We have shown the differential conductance $\frac{\partial I}{\partial V}$ as well for five dots which can be compared with DMRG results for spinless case \cite{Alexander}. 
\section{Results for two levels inside the bias window}
\label{result2}
\begin{figure}
\includegraphics[scale=0.5]{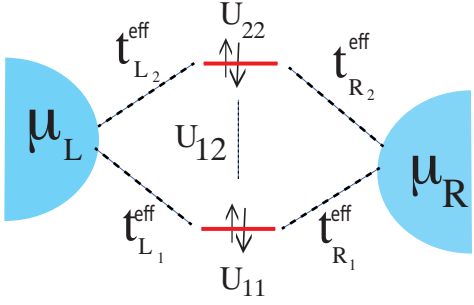}
\caption{\label{fig:proposed-model} Schematical model for a two level system.}
\end{figure}

\begin{figure}
\includegraphics[scale=0.3]{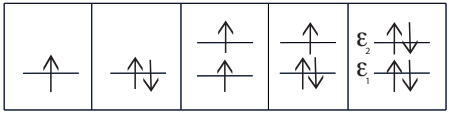}
\caption{\label{fig:5cases} Five proposed configurations to extract the values of $\varepsilon_1$, $\varepsilon_2$, $U_{11}$, $U_{22}$ and $U_{12}$.}
\end{figure}

\begin{figure}
\includegraphics[scale=0.25]{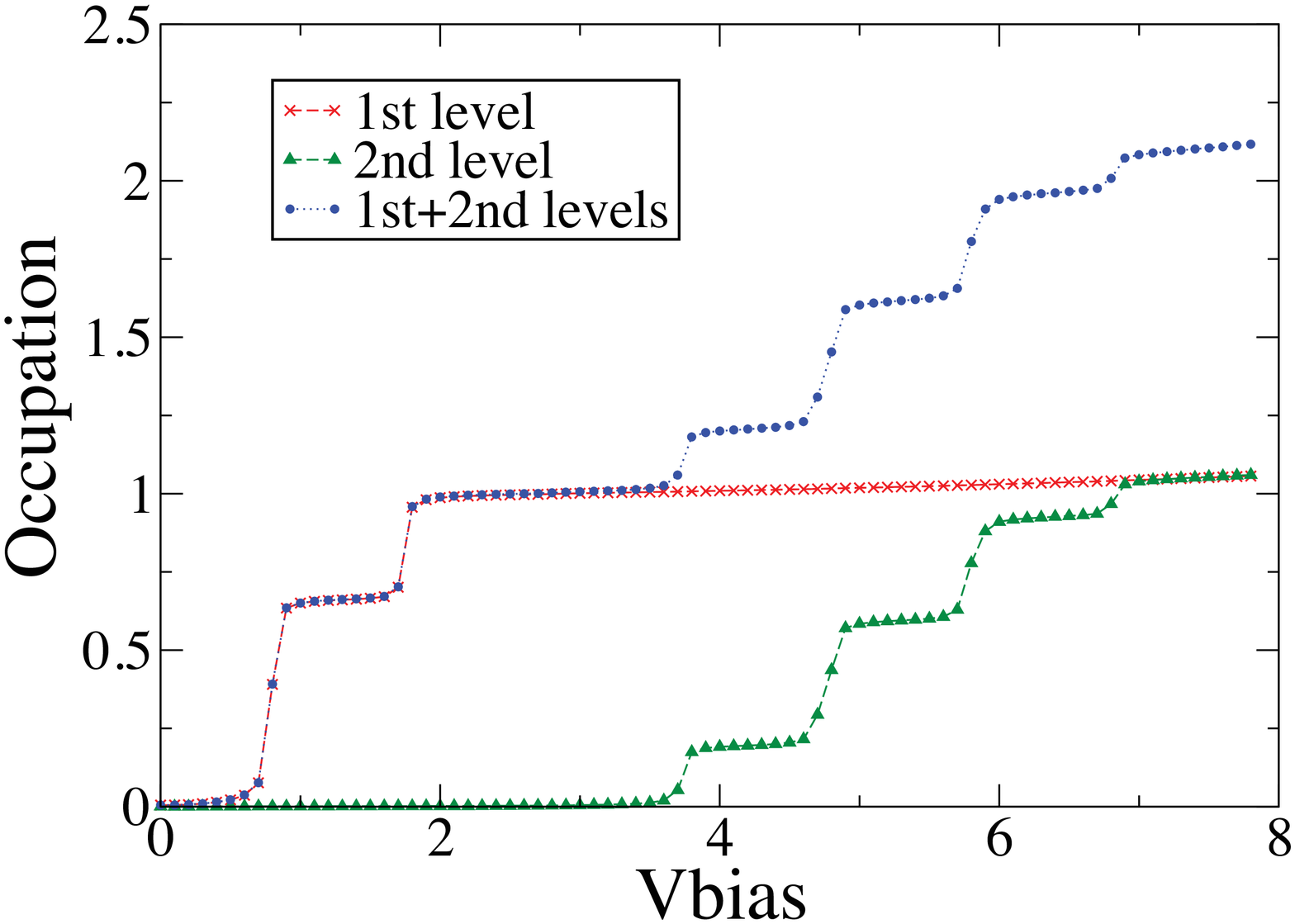}
\caption{\label{fig:n1n2n3n4}Occupation of a two quantum dot chain with $t=1$, $t_c=6$, $t_{L,R}=0.3$, $V_g=1.4$, intial $U=1$ which lead to $\varepsilon_1=0.4$, $\varepsilon_2=1.86$, $U_{11,22}=0.46$, $U_{12}=0.53$. $t^{\text{eff}}_{0 \to 1\uparrow}=t^{\text{eff}}_{1\uparrow \to 1\downarrow}=t^{\text{eff}}_{1\downarrow \to 2\uparrow}=t^{\text{eff}}_{2\uparrow \to 2\downarrow}=0.212132$. The red curve shows the occupation of the first level while the green one shows the occupation of the second level. The blue one is the sum of the occupations. The six steps shown in this curve correspond to different transfer process presented in Fig.~\ref{fig:analyze}.  }
\end{figure}

\begin{figure}
\includegraphics[scale=0.4]{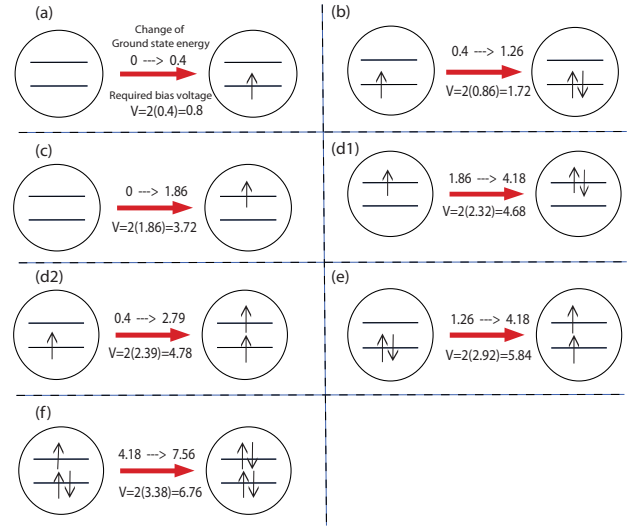}
\caption{\label{fig:analyze} Different transfer process of electrons corresponding to six consecutive steps shown in Fig.~\ref{fig:n1n2n3n4}. (a) shows the transport of the first step in density curve (b) shows the second step and so on. Other processes play a role but they do not show up as separate steps in this graph.}
\end{figure}

\begin{figure}[b]
\includegraphics[scale=0.32]{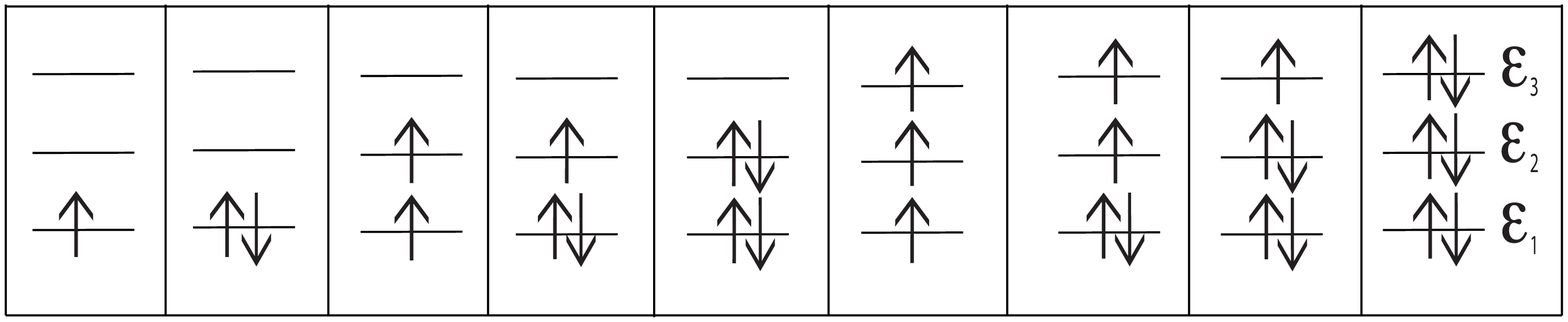}
\caption{\label{fig:9cases} Nine proposed situations to extract the values of $\varepsilon_1$, $\varepsilon_2$,  $\varepsilon_3$, $U_{11,22,33}$ , $U_{12,23,13}$.}
\end{figure}

\begin{figure}[t]
\includegraphics[scale=0.25]{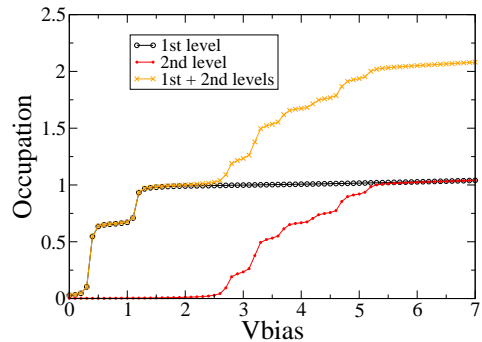}
\caption{\label{fig:3dotsmap}Occupation of a three quantum dot chain with $Vg=1.6$, $U=1.2$, $t_{L,R}=0.4$, $t_c=6$, $t=1$. Extracted effective couplings are $t^{\text{eff}}_{{L,R}_{0 \to 1\uparrow, 1\uparrow \to 1\downarrow}}=0.2$, $t^{\text{eff}}_{{L,R}_{1\downarrow \to 2\uparrow}}= 0.275279 $, $t^{\text{eff}}_{{L,R}_{2\uparrow \to 2\downarrow}}= 0.282842$.}
\end{figure}

For most experimentally relevant situations, the single level problem with interaction will be adequate. However, if the molecule possesses a symmetry (which is not destroyed by an imbalance in the contact geometry) molecular orbitals may become degenerate. Indeed, for chains with more than one site, we find degeneracies in the spectrum of the isolated molecule.\\
In this section we therefore consider a problem with two degenerate levels (see Fig.~\ref{fig:proposed-model}) which may lie inside the bias window. For this case, we do not know of reliable calculations to compare our results with. However, in view of the good agreement with DMRG for the single (interacting) orbitals, we expect the results presented here to be reliable.
We label the four states for two levels $1_{\uparrow}$, $1_{\downarrow}$ , $2_{\uparrow}$, $2_{\downarrow}$ respectively. 
We map our system (shown in Fig.~\ref{fig:model}) to a model with two orbitals with chemical potential that may be degenerate and include the intra-level Coulomb interaction ($U_{11}$ and $U_{22}$) and the inter-level Coulomb interaction ($U_{12}$). We assume that the inter-level interaction does not depend on spin (Fig.~\ref{fig:proposed-model}). In this case we should calculate $\varepsilon_1$, $\varepsilon_2$, $U_{11}$, $U_{22}$ and $U_{12}$. For this purpose, we calculate the ground state energy by FVC parameterization in the five cases shown in Fig.~\ref{fig:5cases}. Thus by having these energy values we can calculate the mentioned energy levels and the Coulomb interactions. From these values, we can then investigate the transport (see Appendix \ref{appx3} for details about the method).\\
An important feature of our method is that it can produce I-V characteristic for rather high bias voltages.\\ 
We start with a chain consisting of two dots. As explained in Sec.~\ref{Method-label}, we first map the spectrum of this chain onto two one-electron levels that can be occupied by spin-up and/or spin-down electrons (see Fig.~\ref{fig:proposed-model}). This is done by calculating the ground state energies for all the configurations shown in Fig.~\ref{fig:5cases}. From these, we find the appropriate $\epsilon$ and $U$ values. The results are shown in Fig.~\ref{fig:n1n2n3n4} for a chain with parameters given in the caption of that figure. The transitions seen in the curve of Fig.~\ref{fig:n1n2n3n4} are displayed in Fig.~\ref{fig:analyze}, together with the predicted energies for these steps based on the parameters of the two-level model of Fig.~\ref{fig:proposed-model}. The first and second steps (a) and (b), will take place when the bias voltage is not high enough to encompass both levels but it is high enough to cross the first level. As this level can be occupied by two electrons we see two steps between $V_b=0$ and $V_b=3$. A higher bias voltage enables the occupation of the second level. The third step shows the addition of an electron to the second level.\\
We also have mapped a Hubbard chain of three interacting dots onto the 3-level model. To extract the nine values of $\varepsilon_1$, $\varepsilon_2$, $\varepsilon_3$, $U_{11,22,33}$ and $U_{12,23,13}$, we considered the nine ground state configurations shown in Fig.~\ref{fig:9cases}. Here we take the two lowest levels to be inside or near the bias window. The result is shown in Fig.~\ref{fig:3dotsmap}. \\
We have implemented our method for at most two levels inside or near the bias window. However, it is possible to use the method for more than two levels inside the bias window which makes the computation time-consuming due to the larger dimension of matrices.\\

\section{Conclusions}
\label{conclu}
In conclusion, we have proposed a method based on DFT which can accurately predict the transport in the weak coupling regime through an interacting chain. In our approach, we map the interacting part of the system to several interacting energy levels and take the Coulomb interactions into account. The dot occupations show different steps corresponding to different transfer processes of electrons from the leads to the interacting region. Our method is a new opening for using DFT first principle calculations to investigate the transport through molecules in the weak and intermediate coupling limit. We do not have the observation of Kondo in our method but it can be included using more advanced approximations. We plan to implement our method into a quantum chemical DFT code to calculate transport through experimentally relevant devices. As ground state DFT can predict excitation energies reasonably well (provided a separate self-consistent calculation
is performed for each particle number and polarization) we should be able to reveal several energy levels. Furthermore the good results for the peak broadenings in our model system are promising, although the coupling strengths in experimental devices suffer from sample to sample variation. The same holds for the dielectric environment which may affect the location of the levels substantially. For this, and for the alignment of the levels to the Fermi energies of the contacts, a constrained DFT approach may be useful.

\section{Acknowledgement}
It is a pleasure to thank K. Capelle, V. Fran\c{c}a and D. Vieira for providing us with the LSDA parameterization for the Hubbard model, P. Schmitteckert for the helpful discussions and M. Leijnse for his useful comments. Financial support was obtained from the EU FP7 program under the grant agreement “SINGLE”.
\appendix
\section{Bethe-Ansatz solution for spinless fermions}
\label{appx1}
Here we briefly describe the numerical approach for finding the exact ground state energy. Takahashi \cite{Takahashi} gives the equations which should be solved for the density $n$, the `quasi-momenta' $k$ and the integer limit $B$. These equations are
\begin{eqnarray}
n=\int_{-B}^{B} \rho(k)  \,dk
\end{eqnarray}	
and
\begin{equation}
1=2\pi \rho(k)-\int_{-B}^{B}T(k,q)\rho(q) \,dq
\end{equation}
where
\begin{equation}
T(k,q)=\frac {\partial }{\partial k}\theta(k,q)
\end{equation}
and
\begin{equation}
\theta(k,q)=2\tan^{-1}\left[  \frac{\frac{U}{2t}\sin(\frac{k-q}{2})} {\cos(\frac{k+q}{2})-\frac{U}{2t}\cos(\frac{k-q}{2})}    \right] 
\end{equation}
$E$ is then given as 
\begin{equation}
E=-\frac{U}{4}-2t\int_{-B}^{B} (\cos k +\frac{U}{2t}) \,dk
\end{equation}
and the exchange-corrolation energy is then defined as 
\begin{equation}
E_{\text{xc}}=E-E(U=0)-U\left(n-\frac{1}{2}\right)^2
\end{equation}
Since the function $\theta$ depends on the momenta, the problem has to be solved self-consistently.
\section{L(S)DA-DFT for the Hubbard model}
\label{appx2}
In order to construct a DFT Hamiltonian with parameteres $U$, $t$ for a Hubbard chain based on L(S)DA, an expression for the exchange-correlation potential is needed. This potential is based on the exact ground state energy.\\
The exact ground state energy $e(n,m,t,U)$ (for density $n=n_{\uparrow}+n_{\downarrow}$ and magnetization $m=n_{\uparrow}-n_{\downarrow}$) of the Hubbard model can be obtained using the Bethe-Ansatz \cite{bethe1, bethe2, bethe3}. An approximate analytical expression for the unpolarized case $(m=0)$ was proposed by K. Capelle \textit{et al.} \cite{Capelle,Capellesame}. This reads
\begin{equation}
e(n\leq 1,m=0,t,U)=-\frac{2t\beta(U/t)}{\pi}\sin \left[ \frac{\pi}{\beta(U/t)}n \right]
\end{equation}
where $n=N/L$ is the Hubbard site occupation, $N$, $L$ are the number of electrons and Hubbard sites respectively and $\beta$ is a function of the ratio $U/t$. It can be determined from the implicit equation
\begin{equation}
-\frac{2t\beta(U/t)}{\pi}\sin(\frac{\pi}{\beta(U/t)})=-4t\int_0^\infty \frac{J_0(x)J_1(x)}{x\left[1+\text{exp}(\frac{xU}{2t})\right]} \,dx
\end{equation}
here $J_{i=0,1}(x)$ are the Bessel functions of the first kind \cite{Capelle, Capellesame}. For $n>1$, the energy is found from the particle-hole symmetry
\begin{equation}
e(n>1,m=0,t,U)=e(2-n,m=0,t,U)+U(n-1)
\end{equation}
From the energy, we can obtain an analytical expression for the exchange-correlation potential which, in the unpolarized case, is
\begin{eqnarray}
&V_{\text{xc}}(n,m=0,t,U)=\frac{\delta e_{\text{xc}}}{\delta n} = \nonumber \\
&\frac{\delta}{\delta n}[ e(n,m=0,t,U)-e(n,m=0,t,0)-e_{H} (n,U)]
\end{eqnarray}
where the Hartree-energy is 
\begin{equation}
e_{H}(n,U)=U n^2 /4
\end{equation}
In the case of non-zero magnetization, the energy expression $e(n,m,t,U)$  has been constructed by V. Fran\c{c}a, D. Vieira and K. Capelle \cite{capelle2}.\\
The linear Hamiltonian matrix dimension for the polarized case with polarization $M$, is ${\binom{L}{(N+M)/2}} {\binom{L}{(N-M)/2}} $, while in the LDA case has the dimension $L$:
\begin{equation}
H=
\begin{pmatrix}
\frac{U}{2}n_1+V_{{\text{xc}}_1} & -t & 0 & \ldots & 0 &0\\
-t & \frac{U}{2}n_2+V_{{\text{xc}}_2} & -t & \ldots& 0  &0 \\
\vdots & \vdots & \vdots & \vdots & \vdots& \vdots\\
0 &0 &0& \ldots& -t & \frac{U}{2}n_{L}+V_{{\text{xc}}_{L}}  
\end{pmatrix}
\end{equation}
This LDA-based Hamiltonian replaces the actual potential felt by an electron when it enters into a site by a time average of electron occupation at that site. In the LSDA Hamiltonian, this potential is a function of the two spin densities ($n_{\uparrow}$, $n_{\downarrow}$). 
\section{Calculating the transport through a Coulomb island}
\label{appx3}
\subsection{Single level inside the bias window}
We start with one level inside the bias window and we explain how the GF for the spinless case and for the case with spin can be derived from the equation of motion (EOM). The time derivative of the $d$ operator (for molecule) and the $c$ operator (for contacts) are (see \cite{Haug})
\begin{equation}
i\dot{{d}_{\alpha}}=\varepsilon_{\alpha}d_{\alpha}+\sum_{\beta \neq \alpha} U_{\alpha\beta}d_{\alpha}n_{\beta}+\sum_{\substack{\eta=L/R \\ q}}t_{\eta k\alpha}^* c_{\eta k \sigma_{\alpha}}
\end{equation}
\begin{equation}
i\dot{c}_{\eta k \sigma_{\alpha}}=\varepsilon_{\eta k}c_{\eta k\alpha}+\sum_{{\alpha}^{\prime}}t_{\eta k\alpha}d_{{\alpha}^{\prime}}.
\end{equation}
Here, $\alpha$ denotes the spin-orbital $\alpha$ and $\sigma_{\alpha}$ is the spin for this $\alpha$. $k$ labels the traveling wave states in the leads $\eta=L,R$ where $L$ and $R$ stand for left and right.
If we consider only one orbital and neglect the spin, the term with $U_{\alpha\beta}$ drops out of the problem. In order to find the current, we use non-equilibrium GF theory, which focuses on the one-particle GF  on the molecule, defined as
\begin{equation}
G_{\alpha\beta}=-i\langle T\{d_{\alpha}(t) d^{\dagger}_{\beta}(t^{\prime})\}\rangle
\end{equation}
where $T$ is the time-ordering operator
\begin{equation}
T\{A(t) B(t^{\prime}) \}=\theta(t-t^{\prime})A(t)B(t^{\prime}) \mp \theta(t^{\prime}-t)B(t^{\prime})A(t)
\end{equation}
$T$ always moves the operators with earlier time argument to the right.\\
After Fourier transformation of the time domain, taking the time ordering carefully into account, the following equation for the GF is found \cite{Haug}:
\begin{equation}
(\omega-\varepsilon_{\alpha})G_{\alpha\beta}(\omega)=\delta_{\alpha\beta}+\sum_{\eta k} t_{\eta} \Gamma_{\eta k}^{\alpha\beta}(\omega)
\end{equation}
where 
\begin{equation}
\Gamma_{\eta k}^{\alpha\beta}(t-t^{\prime})=-i \langle T c_{\eta k \sigma_{\alpha}}(t) d_{\beta}^{\dagger}(t^{\prime}) \rangle
\end{equation}
Using the EOM for $c_{\eta k \sigma_{\alpha}}(t)$, an equation for $\Gamma_{\eta k}^{\alpha\beta}$ is found:
\begin{equation}
(\omega-\varepsilon_{\eta k})\Gamma_{\eta k}^{\alpha\beta}(t-t^{\prime})=t_{\eta} G_{\alpha\beta}(\omega)
\label{e3}
\end{equation}
Using the second equation to eliminate $\Gamma_{\eta k}^{\alpha\beta}$, we arrive at
\begin{equation}
(\omega-\varepsilon_{\alpha}-\Sigma_0(\omega)) G_{\alpha\beta}=\delta_{\alpha\beta}
\end{equation}
where 
\begin{equation}
\Sigma_{0}(\omega)=\sum_{\eta k } \frac{{\vert t_{\eta}\vert}^2}{\omega-\epsilon_{\eta k }}
\end{equation}
is the self-energy. The self-energy has a real (Hermitian) part which has the effect of shifting the resonant energies and which reflects asymmetries of the densities of states near those resonances. The imaginary (non-Hermitian) part broadens the resonances, reflecting the hybridization of the states of the central region with those of the leads.\\
Taking $\alpha=\beta$ and writing $\Sigma_0^r(\omega)=\Lambda(\omega)+i\Gamma /2$, we can write
\begin{equation}
G_{\alpha\alpha}^r(\omega)=\frac {1}{\varepsilon_{\alpha}-\omega-\Lambda-i\Gamma /2}
\end{equation}
which is a Lorentzian function.\\
Next, we consider an interacting dot for spin-1/2 fermions. In that case, the EOM leads to the following equation for the GF:
\begin{equation}
(\omega-\varepsilon_{\alpha}) G_{\alpha\beta}(\omega)=\delta_{\alpha\beta}+\sum_{\gamma \neq \alpha} U_{\alpha\gamma} G_{\alpha\gamma\beta}^{(2)}(\omega)+\sum_{\eta k}t_{\eta} \Gamma_{\eta k}^{\alpha\beta}(\omega)
\label{e1}
\end{equation}
while the equation for $\Gamma_{\eta k}^{\alpha\beta}$ remains the same. We have introduced a new GF $G_{\alpha\gamma\beta}^{(2)}$, which is defined as $G^{(2)}_{\alpha\gamma\beta}=-i\langle T\{d_{\alpha}(t) n_{\gamma}(t) d^{\dagger}_{\beta}(t^{\prime})\}\rangle$ with $n_{\gamma}(t)=d_{\gamma}(t) d_{\gamma}^{\dagger}(t)$. This GF satisfies an EOM:
\begin{eqnarray}
&\displaystyle{(\omega-\varepsilon_{\alpha}-U_{\alpha\gamma}) G^{(2)}_{\alpha\gamma\beta}=\langle n_{\gamma}\rangle \delta_{\alpha\beta}}\nonumber\\
& \displaystyle{+\sum_{\eta k}(  t_{\eta}^* \Gamma_{1,\eta k}^{(2)\alpha\beta}+ t_{\eta} \Gamma_{2,\eta k}^{(2)\alpha\beta}- t_{\eta}^* \Gamma_{3,\eta k}^{(2)\alpha\beta}) }
\label{e2}
\end{eqnarray}
where
\begin{equation}
\Gamma_{1,\eta k}^{(2)\alpha\beta}=-i \langle T\{c_{\eta k\sigma_{\alpha}}(t) n_{\gamma}(t) d^{\dagger}_{\beta}(t^{\prime})\}\rangle
\end{equation}
\begin{equation}
\Gamma_{2,\eta k}^{(2)\alpha\beta}=-i \langle T\{c_{\eta k\sigma_{\gamma}}(t) d_{\alpha}(t) d_{\gamma}(t) d^{\dagger}_{\beta}(t^{\prime})\}\rangle
\end{equation}
\begin{equation}
\Gamma_{3,\eta k}^{(2)\alpha\beta}=-i \langle T\{c_{\eta k\sigma_{\gamma}}(t) d^{\dagger}_{\gamma}(t) d_{\alpha}(t) d^{\dagger}_{\beta}(t^{\prime})\}\rangle
\end{equation}
We now neglect correlation between the central region and the leads by keeping only $\Gamma_{1,\eta k}^{(2)\alpha\beta}$ which we approximate as $\Gamma_{1,\eta k}^{(2)\alpha\beta}=\langle n_{\gamma}\rangle \Gamma_{\eta k}^{\alpha\beta}$. Note that this mean field approximation only concerns the coupling between the central region and the leads, but not the Coulomb correlations within the central region. Thus by substituting $G^{(2)}$ from (\ref{e2}) to (\ref{e1}) and eliminating $\Gamma$ by an equation like (\ref{e3}), it yields 
\begin{eqnarray}
&\displaystyle{ G_{\alpha\alpha}(\omega)=} \nonumber \\
&\displaystyle{ \frac{\omega-\varepsilon_{\alpha}-(1-\langle n_{\beta}\rangle)U}{(\omega-\varepsilon_{\alpha}-U)(\omega-\varepsilon_{\alpha})-\Sigma^{r}[\omega-\varepsilon_{\alpha}-(1-\langle n_{\beta}\rangle)U]} }
\end{eqnarray}
where $\Sigma^{r}=\Sigma^{r}_{L}+\Sigma^{r}_{R}$ and
\begin{equation}
\Sigma^r_j(\omega)= \frac {-t_j^2}{t_c}z_j(\omega)
\label{sigmar}
\end{equation}
and $\text{Im}(z_j)>0$, $z_j=-q_j \pm \sqrt{q_j^2-1}$, $q_j=\frac{\omega-E_F \pm V/2}{2t_c}$ and $E_F$ is the Fermi energy of the grounded lead.\\
To calculate the density self-consistently, the calculation of the lesser GF is also required
\begin{equation}
\langle n_{\alpha}\rangle =\int \frac{G_{\alpha\alpha}^{<}(\omega)}{2\pi i} \,d\omega 
\label{integration} 
\end{equation} 
which can be found from the Keldysh equation Eq (\ref{gless}).
\begin{equation}
G_{\alpha\alpha}^{<}(\omega)=G_{\alpha\alpha}^{r}(\omega) \Sigma_0^{<}(\omega) G_{\alpha\alpha}^{a}(\omega)
\label{gless}
\end{equation}
$G^a$ is the advanced GF and the lesser self-energy is 
\begin{equation}
\Sigma_0^{<}(\omega)=2\sum_{j=L,R} \frac{t_j^2}{t_c}f(\omega,\mu_j) \sqrt{1-\frac{\omega^2}{(4t_c)^2}}
\label{sigma0}
\end{equation}
Once the retarded and advanced GF are known, the current can be calculated from a Landauer type of equation
\begin{equation}
I=\frac{ie}{h}\int \text{Tr}\{ \frac{\Gamma_L\Gamma_R}{\Gamma_L+\Gamma_R}(G^r-G^a)\}(f(\omega,\mu_L)-f(\omega,\mu_R)) \,d\omega
\end{equation}
where $\Gamma_j=i(\Sigma_j^r- \Sigma_j^{r^{\dagger}})$.

\subsection{Two levels inside the bias window}
Here we explain the transport calculation for the case we have two levels inside the bias window. Once we have calculated the energy values and Coulomb interaction by FVC parameterization, we can use a many-body approach which is a generalization of the technique described above. For the one-particle GF, we find the same equation as above:
\begin{eqnarray}
&\displaystyle{(\omega-\varepsilon_{\alpha})G_{\alpha\beta}=\delta_{\alpha\beta}+\sum_{\gamma\neq\alpha} U_{\alpha\gamma}G^{(2)}_{\alpha\gamma\beta}+}\nonumber\\
&\displaystyle{\sum_ {\alpha^{\prime}}\Sigma_{\alpha\alpha^{\prime}} G_{\alpha^{\prime}\beta}}.
\label{set1}
\end{eqnarray}
where $\Sigma_{\alpha\alpha^{\prime}}$ is the self-energy 
\begin{equation}
\Sigma_{\alpha\alpha^{\prime}}=\sum_{\eta k} \frac{t_{\eta k\alpha^{\prime}}t^*_{\eta k\alpha}}{\omega-\epsilon_{\eta k}}
\end{equation}
For $G^{(2)}$ we obtain the EOM
\begin{eqnarray}
&\displaystyle{(\omega-\varepsilon_{\alpha}-U_{\alpha\gamma}) G^{(2)}_{\alpha\gamma\beta}=\langle n_{\gamma}\rangle \delta_{\alpha\beta}}\nonumber\\
& \displaystyle{+\sum_{\delta\neq\alpha,\gamma} {U_{\alpha\delta}G^{(3)}_{\alpha\gamma\delta\beta}} +\sum_ {\alpha^{\prime}} {\Sigma_{\alpha\alpha^{\prime}} G^{(2)}_{\alpha^{\prime}\gamma\beta}}}
\label{set2}
\end{eqnarray}
Eq (\ref{set2}) is not closed as a new GF, $G^{(3)}_{\alpha\gamma\delta\beta}$, is generated in deriving the equation for $G^{(2)}_{\alpha\gamma\beta}$.
The new GF, $G^{(3)}_{\alpha\gamma\delta\beta}$, is
\begin{equation}
G^{(3)}_{\alpha\gamma\delta\beta}=-i\langle T\{d_{\alpha}(t) n_{\gamma}(t) n_{\delta}(t) d^{\dagger}_{\beta}(t^{\prime})\} \rangle
\end{equation}
The EOM for $G^{(3)}_{\alpha\gamma\delta\beta}$ reads
\begin{eqnarray}
&\displaystyle{(\omega-\varepsilon_{\alpha}-U_{\alpha\gamma}-U_{\alpha\delta})G^{(3)}_{\alpha\gamma\delta\beta}=\langle n_{\gamma} n_{\delta} \rangle \delta_{\alpha\beta}}\nonumber\\
& \displaystyle{+\sum_{\epsilon\neq\alpha,\gamma} U_{\alpha\epsilon}G^{(4)}_{\alpha\gamma\delta\epsilon\beta}+ \sum_ {\alpha^{\prime}}\Sigma_{\alpha\alpha^{\prime}} G^{(3)}_{\alpha^{\prime}\gamma\delta\beta}}
\label{set3}
\end{eqnarray}
which introduces another new GF, $G^{(4)}_{\alpha\gamma\delta\epsilon\beta}$
\begin{equation}
G^{(4)}_{\alpha\gamma\delta\epsilon\beta}=-i\langle T\{d_{\alpha}(t) n_{\gamma}(t) n_{\delta}(t) n_{\epsilon}(t) d^{\dagger}_{\beta}(t^{\prime})\} \rangle
\end{equation}
for which the EOM is
\begin{eqnarray}
&\displaystyle{(\omega-\varepsilon_{\alpha}-U_{\alpha\gamma}-U_{\alpha\delta}-U_{\alpha\epsilon}) G^{(4)}_{\alpha\gamma\delta\epsilon\beta}=\langle n_{\gamma} n_{\delta} n_{\epsilon}\rangle \delta_{\alpha\beta} }\nonumber \\
&\displaystyle{ +\sum_{\alpha^{\prime}}{\Sigma_{\alpha\alpha^{\prime}} G^{(4)}_{\alpha^{\prime}\gamma\delta\epsilon\beta}}}
\label{set4}
\end{eqnarray}
At this stage, the process of generating new GF stops, as the EOM for $G^{(4)}$ does not generate higher-order GFs.\\
We now must solve the set of equations (\ref{set1}), (\ref{set2}), (\ref{set3}) and (\ref{set4}) for the GFs $G_{\alpha \beta}$ to $G^{(4)}_{\alpha\gamma\delta\epsilon\beta}$. We organise these GFs into a $340 \times 4$ array 
\begin{equation}
\mathcal{G}_{\Lambda \beta}=(G_{\alpha \beta}, G^{(2)}_{\alpha^{\prime}\gamma\beta}, G^{(3)}_{\alpha^{\prime\prime}\gamma^{\prime}\delta\beta}, G^{(4)}_{\alpha^{\prime\prime\prime}\gamma^{\prime\prime}\delta^{\prime}\epsilon\beta})^T
\end{equation}
As all indices $\alpha,\beta,...$ run over four states, it is easy to see that the first index of this array runs over $4+16+64+256=340$ values. The equation for $\mathcal{G}$ can be written in the form
\begin{equation}
\mathcal{G}_0^{-1} \mathcal{G}= \langle \tilde{n} \rangle+ \Sigma \mathcal{G}
\label{GG}
\end{equation}
Here, $\mathcal{G}_0^{-1}$ is a $340 \times 340$ matrix, which, in the frequency domain assume the form
\begin{widetext}
\begin{equation}
\mathcal{G}_0^{-1}(\omega) =
\begin{pmatrix}
\framebox { \parbox {1.5cm}{ \center{$ \omega-\varepsilon_{\alpha}$} \vspace*{0.2cm}} } & \qquad &  \qquad & \qquad \\
\qquad & \framebox {\parbox{3cm}{ \center{ $\omega-\varepsilon_{\alpha^{\prime}}-U_{\alpha^{\prime}\gamma} $}  \vspace*{0.2cm} } } &\qquad &\qquad\\
\qquad & \qquad & \framebox {\parbox{4cm}{ \center{ $\omega-\varepsilon_{\alpha^{\prime\prime}}-U_{\alpha^{\prime\prime}\gamma^{\prime}}-U_{\alpha^{\prime\prime}\delta} $}  \vspace*{0.2cm}} } & \qquad \\
 &\qquad  &\qquad  &\qquad & \framebox {\parbox{5cm}{ \center{$ \omega-\varepsilon_{\alpha^{\prime\prime\prime}}-U_{\alpha^{\prime\prime\prime}\gamma^{\prime\prime}}-U_{\alpha^{\prime\prime\prime}\delta^{\prime}}-U_{\alpha^{\prime\prime\prime}\epsilon} $}  \vspace*{0.2cm}} }
\end{pmatrix}
\end{equation}
\end{widetext}
and $\langle\tilde{n}\rangle$ is a $340 \times 4$ array
\begin{equation}
(\delta_{\alpha\beta}, \langle n_{\gamma} \rangle \delta_{\alpha^{\prime}\beta},\langle n_{\gamma^{\prime}} n_{\delta} \rangle \delta_{\alpha^{\prime\prime}\beta}, \langle n_{\gamma^{\prime\prime}} n_{\delta\prime} n_{\epsilon} \rangle \delta_{\alpha^{\prime\prime\prime}\beta} )^T
\end{equation}
and $\Sigma$ is the $340 \times 340$ array with elements $\Sigma_{\alpha_{\Lambda},\alpha_{{\Lambda}^{\prime}}}$, where $\alpha_{\Lambda}$ denotes the index $\alpha$ of the composed index $\Lambda=(\alpha, \alpha^{\prime}\gamma, \alpha^{\prime\prime}\gamma^{\prime}\delta, \alpha^{\prime\prime\prime}\gamma^{\prime\prime}\delta^{\prime}\epsilon)$.\\
In order to find the lesser GF $\mathcal{G}^<$, from which $\langle \tilde{n} \rangle$ can be found, we should use a Keldysh or Kadanoff-Baym equation. These equations are conveniently derived from the Langreth rules \cite{Langreth}. These rules apply to the GF $\mathcal{G}$ which is found from Eq (\ref{GG}). Therefore, using the notation of that equation, the Kadanoff-Baym equation can be written as 
\begin{equation}
\mathcal{G}_0^{-1} \mathcal{G}^<= \Sigma^r \mathcal{G}^< + \Sigma^< \mathcal{G}^a .
\label{GG2}
\end{equation}
where $\mathcal{G}^a$ is found as 
\begin{equation}
\mathcal{G}^a = (\mathcal{G}_0^{-1}-\Sigma^a)^{-1} \langle \tilde{n} \rangle .
\label{GG3}
\end{equation}
In electron transport theory, the Keldysh equation,
\begin{equation}
\mathcal{G}^<=  (\langle \tilde{n} \rangle+  \mathcal{G}^r \Sigma^r) \mathcal{G}_0 ^< (\langle \tilde{n} \rangle+ \mathcal{G}^a \Sigma^a)+ \mathcal{G}^r \Sigma^< \mathcal{G}^a .
\end{equation}
is often used, with only the last term on the right hand side, as it can be shown for transport through a single channel, the first term vanishes for the single particle GF $G_{\alpha\beta}$. However, this is not the case when the `higher' GFs $G^{(2)}$ \textit{etc.} are included (this was also pointed out by Song \textit{et al.} \cite{Bo}).
\begin{figure}[t]
\begin{center}
\includegraphics[scale=0.25]{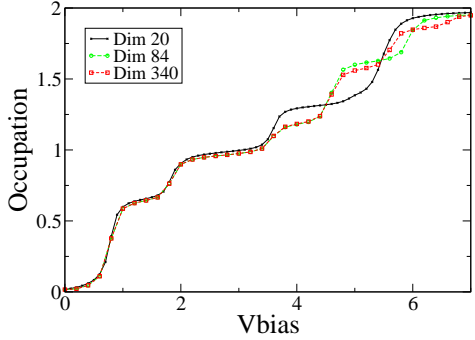}
\caption{\label{fig:3Dims}Occupation of a two quantum dot chain. The chosen parameters are $\varepsilon_1=0.4$, $\varepsilon_2=1.8$, $U_{intra-level}= U_{inter-level}=0.5$, $\Gamma_{L,R}=0.05$, using wide band limit.}
\end{center}
\end{figure}
For the calculation of the integration in Eq (\ref{integration}) one has to calculate the inverse of $G_0$, many times (depending on the number of the integration points and the number of the required iterations to solve the problem self-consistently), making the computation time-consuming. Therefore one could think of using the following approximations which cause reduction of the matrix dimension to $84\times84$ and $20\times20$ respectively.
\begin{eqnarray}
&\displaystyle{G^{(4)}_{\alpha\gamma\delta\epsilon\beta}=-i\langle T\{d_{\alpha}(t) n_{\gamma}(t) n_{\delta}(t) n_{\epsilon}(t) d^{\dagger}_{\beta}(t^{\prime})\}\rangle \simeq}\nonumber\\
&\displaystyle{\frac{1}{3}[\langle n_{\gamma}\rangle G^{(3)}_{\alpha\delta\epsilon\beta}+ \langle n_{\delta}\rangle  G^{(3)}_{\alpha\gamma\epsilon\beta}+ \langle n_{\epsilon}\rangle G^{(3)}_{\alpha\gamma\delta\beta}] }
\end{eqnarray}
\begin{eqnarray}
&\displaystyle{G^{(3)}_{\alpha\gamma\delta\beta}=-i\langle T\{d_{\alpha}(t) n_{\gamma}(t) n_{\delta}(t) d^{\dagger}_{\beta}(t^{\prime})\}\rangle \simeq}\nonumber\\
&\displaystyle{\frac{1}{2} [\langle n_{\delta}\rangle  G^{(2)}_{\alpha\gamma\beta}+ \langle n_{\gamma} \rangle G^{(2)}_{\alpha\delta\beta}]}
\end{eqnarray}
Fig.~\ref{fig:3Dims} shows the effect of the approximations on the occupation. The curve corresponding to dimension $20\times20$ shows four steps in agreement with the full GF while the one based on dimension $84\times84$ depicts five correct steps and finally six steps has been gained from the exact solution (dimension $340\times340$) as we discussed. However, for the lower biases the results based on approximations are still valid. \\
As we can see in Fig.~\ref{fig:3Dims} the density does not exceed 2 while in Fig.~\ref{fig:n1n2n3n4} the occupation exceeds 2 and this can be explained by the difference between the self-energies used in these two figures. The wide band limit has been used in Fig.~\ref{fig:3Dims} which supplies constant self-energies while in Fig.~\ref{fig:n1n2n3n4} the self-energy reflects the density of states in the leads not being constant. Therefore, at different biases, the leads supply a different number of electrons.
\bibliography{biblio}
\end{document}